\documentstyle[e-e-ijmpa,psfig,equations]{article}

\renewcommand{\thefootnote}{\fnsymbol{footnote}}
\newenvironment{Eqnarray}%
     {\arraycolsep 0.14em\begin{eqnarray}}{\end{eqnarray}}

\begin{document}
\def\ie{{\it i.e.}}
\def\etal{{\it et al.}}
\def\phm{\phantom{-}}
\def\eg{{\it e.g.}}
\def\ifmath#1{\relax\ifmmode #1\else $#1$\fi}
\def\ls#1{\ifmath{_{\lower1.5pt\hbox{$\scriptstyle #1$}}}}
\def\calv{{\cal V}}
\def\beqno{\begin{eqalignno}}
\def\eeqno{\end{eqalignno}}
\def\beq{\begin{equation}}
\def\eeq{\end{equation}}
\def\beqa{\begin{Eqnarray}}
\def\eeqa{\end{Eqnarray}}
\def\half{\ifmath{{\textstyle{1 \over 2}}}}
\def\fourth{\ifmath{{\textstyle{1 \over 4}}}}
\def\hsm{h\ls{\rm SM}^0}
\def\mhsm{m_{\hsm}}
\def\hl{h^0}
\def\mhl{m_{\hl}}
\def\hh{H^0}
\def\mhh{m_{\hh}}
\def\hpm{H^{\pm}}
\def\mhpm{m_{\hpm}}
\def\ha{A^0}
\def\mha{m_{\ha}}
\def\tanb{\tan\beta}
\def\sinb{\sin\beta}
\def\cosb{\cos\beta}
\def\cosbma{\cos(\beta-\alpha)}
\def\sinbma{\sin(\beta-\alpha)}
\def\costwobma{\cos^2(\beta-\alpha)}
\def\sintwobma{\sin^2(\beta-\alpha)}
\def\cosbpa{\cos(\beta+\alpha)}
\def\sinbpa{\sin(\beta+\alpha)}
\def\mt{m_t}
\def\mw{m\ls W}
\def\mz{m\ls Z}
\def\mzz{m\ls{Z}^2}
\def\mww{m\ls{W}^2}
\def\eqs#1#2{eqs.~(\ref{#1}) and (\ref{#2})}
\def\eq#1{eq.~(\ref{#1})}
\def\Eqs#1#2{Eqs.~(\ref{#1}) and (\ref{#2})}
\def\Eq#1{Eq.~(\ref{#1})}
\def\Ref#1{Ref.~\citenum{#1}}
\def\Refs#1#2{Refs.~\citenum{#1} and \citenum{#2}}

\def\lsim{\mathrel{\raise.3ex\hbox{$<$\kern-.75em\lower1ex\hbox{$\sim$}}}}
\def\gsim{\mathrel{\raise.3ex\hbox{$>$\kern-.75em\lower1ex\hbox{$\sim$}}}}

\def\ppnp#1#2#3{{\sl Prog. Part. Nucl. Phys. }{\bf #1} (#2) #3}
\def\npb#1#2#3{{\sl Nucl. Phys. }{\bf B#1} (#2) #3}
\def\plb#1#2#3{{\sl Phys. Lett. }{\bf B#1} (#2) #3}
\def\prd#1#2#3{{\sl Phys. Rev. }{\bf D#1} (#2) #3}
\def\prl#1#2#3{{\sl Phys. Rev. Lett. }{\bf #1} (#2) #3}
\def\prc#1#2#3{{\sl Phys. Reports }{\bf #1} (#2) #3}
\def\zpc#1#2#3{{\sl Z. Phys. }{\bf C#1} (#2) #3}
\def\ptp#1#2#3{{\sl Prog.~Theor.~Phys.~}{\bf #1} (#2) #3}
\def\aop#1#2#3{{\sl Ann.~of~Phys.~}{\bf #1} (#2) #3}
%
%
\def\PRL#1&#2&#3&{\sl Phys.\ Rev.\ Lett.\ \bf #1\rm ,\ #2\ (19#3)}
\def\PRB#1&#2&#3&{\sl Phys.\ Rev.\ \bf #1\rm ,\ #2\ (19#3)}
\def\PRP#1&#2&#3&{\sl Phys.\ Rep.\ \bf #1\rm ,\ #2\ (19#3)}
\def\NPB#1&#2&#3&{\sl Nucl.\ Phys.\ \bf #1\rm ,\ #2\ (19#3)}
\def\PL#1&#2&#3&{\sl Phys.\ Lett.\ \bf #1\rm ,\ #2\ (19#3)}
\def\ZP#1&#2&#3&{\sl Z.\ Phys.\ \bf #1\rm ,\ #2\ (19#3)}
%
\def\NPRL#1&#2&#3&{\sl Phys.\ Rev.\ Lett.\ \bf #1\ \rm (19#2)\ #3}
\def\NPR#1&#2&#3&{\sl Phys.\ Rev.\ \bf #1\ \rm (19#2)\ #3}
\def\NNP#1&#2&#3&{\sl Nucl.\ Phys.\ \bf #1\ \rm (19#2)\ #3}
\def\NPL#1&#2&#3&{\sl Phys.\ Lett.\ \bf #1\ \rm (19#2)\ #3}
\def\NZP#1&#2&#3&{\sl Z.\ Phys.\ \bf #1\ \rm (19#2)\ #3}
%
%
\noindent
\rightline{SCIPP 98/10}
\rightline{March 1998}
\rightline{hep-ph/9806331}
\vspace{2.5cm}

\thispagestyle{empty}
\centerline{{\Large\bf Probing the MSSM Higgs Sector at 
an \boldmath$e^-e^-$ Collider}}
\vskip1pc
\begin{center}
{\large Howard E. Haber}
\vskip3pt
{\it Santa Cruz Institute for Particle Physics \\
University of California, Santa Cruz, CA 95064, USA}
     \\[2cm]

Abstract \\
\end{center}
The theoretical structure of the Higgs sector of the Minimal
Supersymmetric Standard Model (MSSM) is briefly described.
An outline of Higgs phenomenology at future lepton colliders
is presented, and some opportunities for probing the
physics of the MSSM Higgs sector at an $e^-e^-$ collider are
considered.
\vfill

\begin{center}
To appear in \\
{\it $e^-e^-$ 97: Proceedings of Electron--Electron Linear
Collider Workshop}\\
Santa Cruz, CA, September 22--24, 1997      \\
{\it International Journal of Modern Physics A}\\
Special Proceedings Issue, June 1998
\end{center}

\clearpage

\setcounter{page}{1}
\normalsize\textlineskip
\centerline{}

\title{PROBING THE MSSM HIGGS SECTOR AT AN \boldmath$e^-e^-$ COLLIDER}

\author{HOWARD E. HABER%
\footnote{Work supported in part by the Department of Energy,
Contract DE-FG03 92ER40689}
}

\address{Santa Cruz Institute for Particle Physics \\
University of California, Santa Cruz, CA 95064, USA}

\maketitle\abstracts{
The theoretical structure of the Higgs sector of the Minimal
Supersymmetric Standard Model (MSSM) is briefly described.
An outline of Higgs phenomenology at future lepton colliders
is presented, and some opportunities for probing the
physics of the MSSM Higgs sector at an $e^-e^-$ collider are
considered.
}

\setcounter{footnote}{0}

\renewcommand{\thefootnote}{\alph{footnote}}

\vspace*{1pt}\textlineskip
\section{Review of the MSSM Higgs Sector}

The Minimal Supersymmetric extension of the Standard Model (MSSM)
contains the Standard Model particle spectrum and the corresponding
supersymmetric partners.\cite{susyrev,susyrevtwo,hehtasi}
In addition, the MSSM must possess two Higgs doublets in
order to give masses to up and down type fermions in a manner consistent
with supersymmetry (and to avoid gauge anomalies introduced by
the fermionic superpartners of the Higgs bosons).
In particular, the MSSM Higgs sector is a CP-conserving
two-Higgs-doublet model, which can be parameterized at tree-level
in terms of
two Higgs sector parameters. This structure arises
due to constraints imposed by
supersymmetry that fix the Higgs quartic couplings in terms
of electroweak gauge coupling constants.

Let $\Phi_1$ and
$\Phi_2$ denote two complex $Y=1$, SU(2)$\ls{L}$ doublet scalar fields.
The most general gauge invariant CP-conserving scalar potential is given by

\vbox{%
\beqno
\calv&=m_{11}^2\Phi_1^\dagger\Phi_1+m_{22}^2\Phi_2^\dagger\Phi_2
-[m_{12}^2\Phi_1^\dagger\Phi_2+{\rm h.c.}]\nonumber\\[6pt]
&\quad +\half\lambda_1(\Phi_1^\dagger\Phi_1)^2
+\half\lambda_2(\Phi_2^\dagger\Phi_2)^2
+\lambda_3(\Phi_1^\dagger\Phi_1)(\Phi_2^\dagger\Phi_2)
+\lambda_4(\Phi_1^\dagger\Phi_2)(\Phi_2^\dagger\Phi_1)
\nonumber\\[6pt]
&\quad +\left\{\half\lambda_5(\Phi_1^\dagger\Phi_2)^2
+\big[\lambda_6(\Phi_1^\dagger\Phi_1)
+\lambda_7(\Phi_2^\dagger\Phi_2)\big]
\Phi_1^\dagger\Phi_2+{\rm h.c.}\right\}\,, \label{pot}
\eeqno }

\noindent
where all parameters introduced above are real.  Supersymmetry imposes
the following constraints on the tree-level values of the quartic
couplings $\lambda_i$:
\beqa \label{bndfr}
&&\lambda_1 =\lambda_2 = \fourth (g^2+g'^2)\,, \qquad\qquad\,
\lambda_3 =\fourth (g^2-g'^2)\,,    \nonumber\\
&&\lambda_4 =-\half g^2\,, \qquad\qquad\qquad\qquad\quad
\lambda_5 =\lambda_6=\lambda_7=0\,.
\eeqa
The tree-level Higgs masses and mixing can now be computed in the
usual manner.\cite{hehtasi90}   The scalar fields will
develop non-zero vacuum expectation values if the mass matrix
$m_{ij}^2$ has at least one negative eigenvalue. Imposing CP invariance
and U(1)$\ls{\rm EM}$ gauge symmetry, the minimum of the potential is
\beq
\langle \Phi_1 \rangle={1\over\sqrt{2}} \left(
\begin{array}{c} 0\\ v_1\end{array}\right), \qquad \langle
\Phi_2\rangle=
{1\over\sqrt{2}}\left(\begin{array}{c}0\\ v_2
\end{array}\right)\,,\label{potmin}
\eeq
where the $v_i$ are real.  It is convenient to introduce
the following notation:
\beq
v^2\equiv v_1^2+v_2^2={4\mw^2\over g^2}=(246~{\rm GeV})^2\,,
\qquad\qquad\tanb\equiv{v_2\over v_1}\,.\label{tanbdef}
\eeq

Of the original eight scalar degrees of freedom, three Goldstone
bosons ($G^\pm$ and $G^0$)
are absorbed and become the longitudinal components of
the $W^\pm$ and $Z$.  The remaining
five physical Higgs particles are: two CP-even scalars ($\hl$ and
$\hh$, with $\mhl\leq \mhh$), one CP-odd scalar ($\ha$) and a charged
Higgs pair ($\hpm$). The squared-mass parameters $m_{11}^2$ and
$m_{22}^2$
can be eliminated by minimizing the scalar potential.  The resulting
squared masses for the CP-odd and charged Higgs states are
\beqa
\mha^2 &=& m_{12}^2(\tan\beta+\cot\beta)\,,\nonumber\\
\mhpm^2 &=& \mha^2+\mw^2\,,
\label{susymhpm}
\eeqa

\noindent and the tree-level neutral CP-even mass matrix is given by
\beq
{\cal M}_0^2 =    \left(
\begin{array}{ll}
\phm\mha^2 \sin^2\beta + m^2_Z \cos^2\beta&
          -(\mha^2+m^2_Z)\sin\beta\cos\beta\\
 -(\mha^2+m^2_Z)\sin\beta\cos\beta&
 \phm\mha^2\cos^2\beta+ m^2_Z \sin^2\beta\end{array}\right)\,.\label{kv}
\eeq
The CP-even Higgs mass eigenstates are linear combinations of
the real components of the neutral Higgs fields
\beqa\label{eigenstates}
\hh&=&\phm(\sqrt{2}\,{\rm Re}~\Phi_1^0-v_1)\cos\alpha+
(\sqrt{2}\,{\rm Re}~\Phi_2^0-v_2)\sin\alpha \nonumber \\
\hl&=&-(\sqrt{2}\,{\rm Re}~\Phi_1^0-v_1)\sin\alpha+
(\sqrt{2}\,{\rm Re}~\Phi_2^0-v_2)\cos\alpha\,,
\eeqa
where $\alpha$ is the angle that diagonalizes the squared-mass matrix.
The eigenvalues of ${\cal M}_0^2$ are
the squared masses of the two CP-even Higgs scalars
\beq
  m^2_{H^0,h^0} = \half \left( \mha^2 + m^2_Z \pm
                  \sqrt{(\mha^2+m^2_Z)^2 - 4m^2_Z \mha^2 \cos^2 2\beta}
                  \; \right)\,.\label{kviii}
\eeq
and the CP-even Higgs mixing angle $\alpha$ is determined by
\beq
  \cos 2\alpha = -\cos 2\beta \left( {\mha^2-m^2_Z \over
                  m^2_{H^0}-m^2_{h^0}}\right)\,,\qquad
  \sin 2\alpha = -\sin 2\beta \left( m^2_{H^0} + m^2_{h^0} \over
                   m^2_{H^0}-m^2_{h^0} \right)\,.\label{kix}
\eeq
From these results, it is easy to obtain:
\beq
\cos^2(\beta-\alpha)={\mhl^2(\mz^2-\mhl^2)\over
\mha^2(\mhh^2-\mhl^2)}\,.
\label{cbmasq}
\eeq
Thus, in the MSSM, two parameters (conveniently chosen to be $\mha$
and $\tanb$) suffice to fix all other tree-level Higgs sector parameters.

\Eq{kviii} implies the following tree-level sum rule:
\beq
\mhl^2+\mhh^2=\mha^2+\mz^2\,.
\eeq
In addition, a
number of important mass inequalities can be derived
from the expressions for the tree-level Higgs masses obtained above:

\vbox{%
\beqno
  m_{h^0} &\leq \mha|\cos 2\beta | \leq \mha \,,\nonumber \\
  m_{h^0} &\leq \mz|\cos 2\beta | \leq \mz \,, \nonumber\\
  m_{H^0} &\geq (\mha^2\sin^2 2\beta+\mz^2)^{1/2}\geq\mz\,, \nonumber\\
  m_{H^0} &\geq (\mz^2\sin^2 2\beta+\mha^2)^{1/2}\geq\mha\,, \nonumber\\
  m_{H^\pm} &\geq\mw\,. \label{kx}
\eeqno
}
Of particular note is the tree-level upper bound of
the mass of the lightest CP-even Higgs boson,
$\mhl\leq \mz|\cos 2\beta|$.
If this prediction were exact, it would imply that the Higgs boson
must be discovered at the LEP-2 collider during its 1998 run, assuming
the expected center-of-mass energy of 190~GeV is achieved
with an integrated luminosity of
150~${\rm pb}^{-1}$.  Absence of a Higgs boson lighter than $\mz$
would then (apparently)
rule out the MSSM.  However, when radiative corrections are included,
the light Higgs mass upper bound may increase significantly.  In the
one-loop leading logarithmic approximation (assuming
$\mha\gsim\mz$),\cite{hhprl}
\begin{equation} \label{mhlapprox}
\mhl^2\lsim\mzz\cos^2\beta+{3g^2 m_t^4\over
8\pi^2\mww}\,\ln\left({M^2_{\tilde t}\over \mt^2}\right)\,,
\end{equation}
where $M_{\tilde t}$ is the (approximate) common mass of the
top-squarks.  Observe that the Higgs mass upper bound is very sensitive
to the value of the top mass and depends logarithmically on the
top-squark masses.  Although eq.~(\ref{mhlapprox}) provides a rough
guide to the Higgs mass upper bound, it is not sufficiently precise for
LEP-2 phenomenology, whose Higgs mass reach depends delicately on the
MSSM parameters.  In addition, in order to perform precision Higgs
measurements and make comparisons with theory, more accurate predictions
for the Higgs sector masses are required.
The formula for the full one-loop
radiative corrected Higgs mass has been obtained in the literature,
although it
is very complicated since it depends in detail on the virtual
contributions of the MSSM spectrum.\cite{honeloop}
Moreover, if the supersymmetry breaking scale is larger than a few
hundred GeV, then renormalization group
(RG) methods are essential for summing up the effects of
large logarithms and obtaining an accurate prediction.

The computation of the RG-improved
one-loop corrections requires numerical integration of a coupled set of
renormalization group
equations.\cite{llog}  (The dominant two-loop next-to-leading
logarithmic results are also known.\cite{hempfhoang})
Although this program has been
carried out in the literature, the procedure is unwieldy
and not easily amenable to large-scale Monte-Carlo analyses.
In \Refs{hhh}{carena}, a simple analytic
procedure for accurately approximating $m_{h^0}$ has been presented.
This method can be easily implemented, and incorporate both the
leading one-loop and two-loop effects and the RG-improvement.
Also included are the leading effects at one loop of supersymmetric
thresholds (the most important effects of this type
arise from third-generation squark mixing).
Here, I shall simply quote two specific bounds, assuming
$\mt=175$~GeV and $M_{\tilde t}\lsim 1$~TeV:
$\mhl\lsim 112$~GeV if top-squark mixing is negligible, while
$\mhl\lsim 125$~GeV if top-squark mixing is ``maximal''.
%
%
Maximal mixing corresponds to
an off-diagonal squark squared-mass that produces the largest value of
$\mhl$.  This mixing leads to an extremely large splitting of top-squark
mass eigenstates.
Current state-of-the-art calculations can obtain a mass bound
for the light CP-even Higgs boson of the MSSM that
is reliable to within a few GeV.

Radiative corrections also modify the tree-level
prediction for the masses of $\hh$ and $\hpm$. However, the
corrections are always smaller than the corrections to
$\mhl$ discussed above.\footnote{
More precisely, the corrections to $\mhh$ are small if $\mha\gsim\mz$.
Conversely, if $\mha\lsim\mz$, then the one-loop
correction to $\mhl^2$ is small, while the correction to $\mhh^2$
is proportional to $g^2(\mt^4/\mw^2)\ln(M^2_{\tilde t}/\mt^2)$.
In contrast, the
leading logarithmically-enhanced correction to $\mhpm^2$ is
proportional to $g^2 m_t^2\ln(M^2_{\tilde t}/\mt^2)$,
independent of the value of $\mha$.}~~It
should be noted that $\mhpm\geq\mw$ is
also not a strict bound when the one-loop corrections are included,
although this tree-level bound does hold
approximately over most of MSSM parameter space (and can be significantly
violated only when $\mha\lsim\mz$ and $\tanb$ is well below 1, a region
of parameter space that is theoretically and
phenomenologically disfavored).

Having reviewed the dependence of the MSSM Higgs masses on the
model parameters, I now turn to the tree-level Higgs couplings.
The Higgs couplings to gauge bosons follow from
gauge invariance and are thus model independent.  For example,
the couplings of the two CP-even Higgs bosons to $W$ and $Z$ pairs
are given by
\beqno
 g\ls{\hl VV}&=g\ls{V} m\ls{V}\sinbma\,, \nonumber \\[3pt]
           g\ls{\hh VV}&=g\ls{V} m\ls{V}\cosbma\,,\label{vvcoup}
\eeqno
where
\beq
g\ls V\equiv\begin{cases}
g,& $V=W\,$,\\ g/\cos\theta_W,& $V=Z\,$. \end{cases}
\label{hix}
\eeq
There are no tree-level couplings of $\ha$ or $\hpm$ to $VV$.
Gauge invariance also determines the strength of the trilinear
couplings of one gauge boson to two Higgs bosons.  For example,
\beqno g\ls{\hl\ha Z}&={g\cosbma\over 2\cos\theta_W}\,,\nonumber \\[3pt]
           g\ls{\hh\ha Z}&={-g\sinbma\over 2\cos\theta_W}\,.
           \label{hvcoup}
\eeqno

In the examples shown above, some of the couplings can be suppressed
if either $\sin(\beta-\alpha)$ or $\cos(\beta-\alpha)$ is very small.
Note that all the vector boson--Higgs boson couplings
cannot vanish simultaneously. From the expressions above, we see
that the following tree-level sum rules must hold separately for $V=W$
and $Z$:
\beqno
g_{\hh V V}^2 + g_{\hl V V}^2 &=
                      g\ls{V}^2m\ls{V}^2\,,\nonumber \\[3pt]
g_{\hl\ha Z}^2+g_{\hh\ha Z}^2&=
             {g^2\over 4\cos^2\theta_W}\,.
\label{sumruletwo}
\eeqno
These results are a
consequence of the tree-unitarity of the electroweak theory.\cite{ghw}
Moreover, if we focus on a given CP-even Higgs state, we note that
its couplings to $VV$ and $\ha V$ cannot be simultaneously
suppressed, since eqs.~(\ref{vvcoup})--(\ref{hvcoup}) imply
that\cite{cpvcoup}
\beq
  g^2_{H ZZ} + 4m^2_Z g^2_{HA^0Z} = {g^2m^2_Z\over
        \cos^2\theta_W}\,,\label{hxi}
\eeq
for $H=\hl$ or $\hh$.  Similar considerations also hold for
the coupling of $\hl$ and $\hh$ to $W^\pm H^\mp$.  We can summarize
the above results by noting that the coupling of $\hl$ and $\hh$ to
vector boson pairs or vector--scalar boson final states is proportional
to either $\sin(\beta-\alpha)$ or $\cos(\beta-\alpha)$ as indicated
below.\cite{hhg,hhgsusy}
\beq
\renewcommand{\arraycolsep}{2cm}
\let\us=\underline
\begin{array}{ll}
  \us{\cos(\beta-\alpha)}&  \us{\sin(\beta-\alpha)}\\[3pt]
       \hh W^+W^-&        \hl W^+W^- \\
       \hh ZZ&            \hl ZZ \\
       Z\ha\hl&          Z\ha\hh \\
       W^\pm H^\mp\hl&  W^\pm H^\mp\hh \\
       ZW^\pm H^\mp\hl&  ZW^\pm H^\mp\hh \\
       \gamma W^\pm H^\mp\hl&  \gamma W^\pm H^\mp\hh
\end{array}
\label{littletable}
\eeq
Note in particular that {\it all} vertices
in the theory that contain at least
one vector boson and {\it exactly one} of the non-minimal Higgs boson
states ($\hh$, $\ha$ or $\hpm$) are proportional to
$\cos(\beta-\alpha)$.

In contrast to the above results, none of the
Higgs self-couplings and Higgs-fermion couplings vanish if either
$\sin(\beta-\alpha)=0$ or $\cos(\beta-\alpha)=0$.
The three-point and four-point Higgs self-couplings depend on the
parameters of the two-Higgs-doublet potential [eq.~(\ref{pot})].
In the MSSM, the tree-level four-point self-couplings are fixed in terms
of the electroweak gauge couplings by virtue of \eq{bndfr}.  The
three-point
couplings are dimensionful, so they depend additionally on $v$ and
$\tanb$.  A complete list of these couplings can be found in the
appendices of \Ref{hhg}.

Supersymmetry imposes a particular structure on the coupling of the
Higgs bosons to the fermions.  In particular, one Higgs doublet
(before symmetry
breaking) couples exclusively to down-type fermions and the other
Higgs doublet couples exclusively to up-type
fermions.  As a result, the charged Higgs boson coupling to fermion pairs
(with all particles pointing into the vertex) is given by
\beq
g_{H^- t\bar b}={g\over{2\sqrt{2}\mw}}\
[m_t\cot\beta\,(1+\gamma_5)+m_b\tan\beta\,(1-\gamma_5)]\,,
\label{hpmqq}
\eeq
where the notation of the 3rd family has been employed, and the effects
of intergenerational mixing (which depends in the same way on the
Cabibbo-Kobayashi-Maskawa mixing matrix as the usual weak charged
current couplings) have been omitted.
The neutral Higgs bosons couplings to $f\bar f$ are flavor diagonal,
and are listed below using 3rd family notation,
relative to the Standard Model value, $gm_f/2\mw$:
\begin{eqaligntwo} \label{qqcouplings}
\hl b\bar b:&~~ -{\sin\alpha\over\cos\beta}=\sin(\beta-\alpha)
-\tan\beta\cos(\beta-\alpha)\,,\nonumber\\[3pt]
\hl t\bar t:&~~~ \phm{\cos\alpha\over\sin\beta}=\sin(\beta-\alpha)
+\cot\beta\cos(\beta-\alpha)\,,\nonumber\\[3pt]
\hh b\bar b:&~~~ \phm{\cos\alpha\over\cos\beta}=\cos(\beta-\alpha)
+\tan\beta\sin(\beta-\alpha)\,,\nonumber\\[3pt]
\hh t\bar t:&~~~ \phm{\sin\alpha\over\sin\beta}=\cos(\beta-\alpha)
-\cot\beta\sin(\beta-\alpha)\,,\nonumber\\[3pt]
\ha b \bar b:&~~~\phm\gamma_5\,{\tan\beta}\,,\nonumber\\[3pt]
\ha t \bar t:&~~~\phm\gamma_5\,{\cot\beta}\,,
\end{eqaligntwo}
where the $\gamma_5$ indicates a pseudoscalar coupling.

All tree-level couplings discussed above are modified by
radiative corrections.  In a first approximation, one can consider the
modifications to the couplings that arise through the renormalization
of the CP-even Higgs mixing angle $\alpha$.  This simply requires a computation
of the radiatively-corrected CP-even Higgs squared-mass matrix.
By diagonalizing this matrix, one extracts the
radiatively corrected value of $\alpha$.\cite{diazmix}
The parameter $\tanb$ is by
assumption an input parameter.\footnote{This is a somewhat subtle
issue, since one must define the input parameters of the theory
through some suitable physical process.  Perhaps the simplest procedure is to
work with a minimally-subtracted definition for $\tanb$.  Then, one can
express all observables in terms of this definition.  Eventually, a
global fit to the observables of the Higgs and higgsino sectors can be
used to experimentally
determine the value of $\tanb$.}~~The most significant implication of
this procedure is the renormalization of $\cos(\beta-\alpha)$,
shown in Fig.~1, which governs
many of the couplings listed above.  However, this approximation clearly
misses some of the radiative corrections that arise from the
radiatively corrected three-point (and four-point) vertex functions.
In most cases, these corrections are not expected to be large,
although a full set of Higgs coupling radiative corrections will be
required to analyze future Higgs precision measurements.

\begin{figure}[h]
\centering
\centerline{\psfig{file=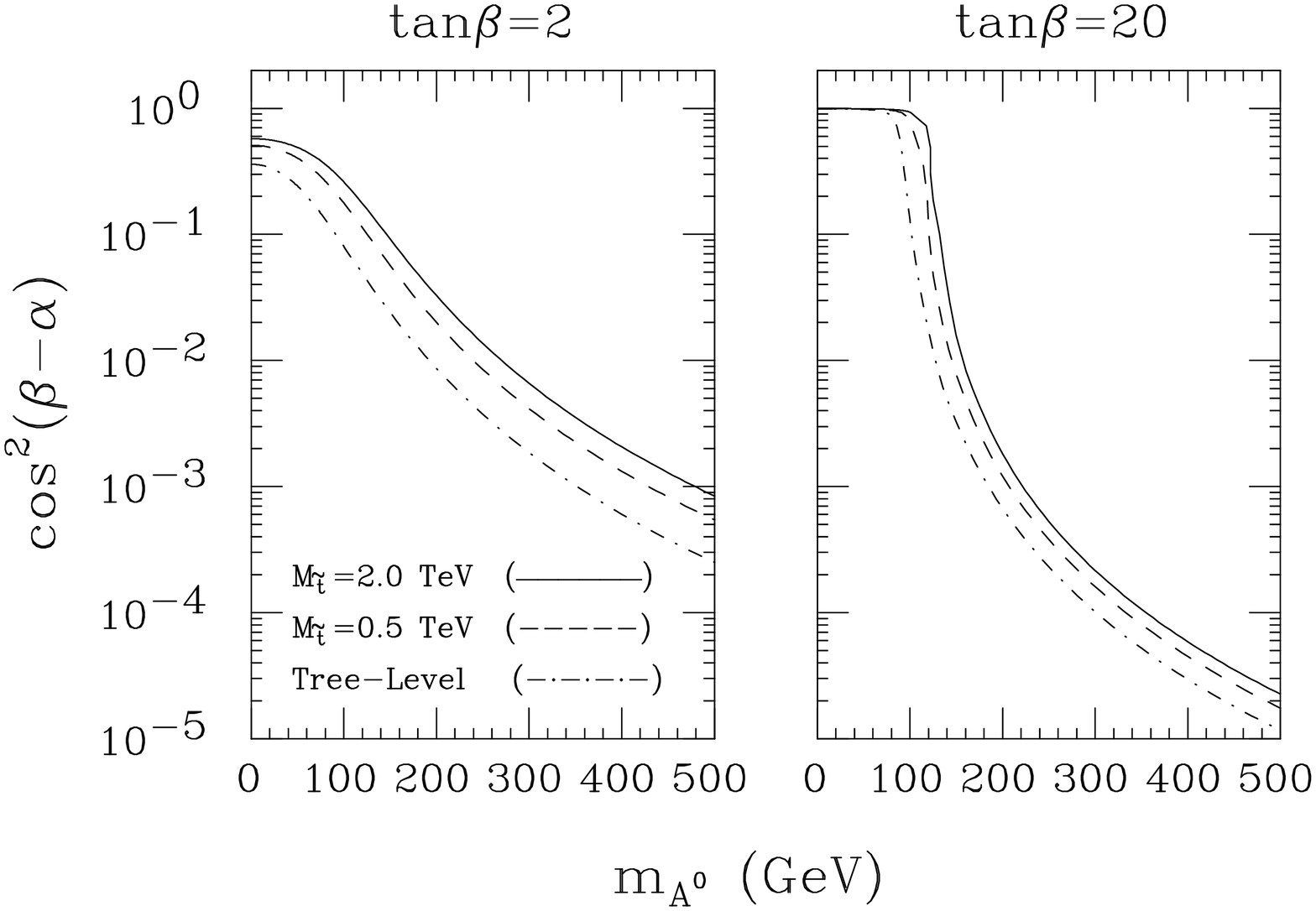,height=7cm,angle=90}}
\fcaption{The value of $\cos^2(\beta-\alpha)$
is shown as a function of
$m_{A^0}$ and $\tan\beta$.
The tree-level prediction is depicted by the dot-dashed line.
The solid and dashed lines include the effects of the
renormalization-group-improved
radiatively-corrected CP-even Higgs squared mass matrix
for two different choices of the (common) top-squark mass, using the
formulae of \Ref{hhh}.  The impact of the radiative corrections grows
with increasing top-squark mass.  Squark mixing effects are neglected;
although they have a negligible effect on the results shown.
}
\end{figure}

\section{The Decoupling Limit}

The pattern of tree-level couplings exhibited in Section 1 can be
understood in the
context of the {\it decoupling limit} of the two-Higgs-doublet model.%
\cite{habernir,DECP}
First, consider the Standard Model Higgs boson ($\hsm$).  At tree-level,
the Higgs quartic self-coupling is related to its mass:
$\lambda=\mhsm^2/v^2$.
This means that one cannot take $\mhsm$ arbitrarily large without the
attendant growth in $\lambda$.  That is, the heavy Higgs limit in the
Standard Model exhibits non-decoupling.  In models of a non-minimal
Higgs sector, the situation is more complex. In some models, it is not
possible to take any Higgs mass much larger than ${\cal O}(v)$ without
finding at least one strong Higgs self-coupling.
In models such as the MSSM, one
finds that the non-minimal Higgs boson masses can be taken large at
fixed Higgs self-couplings.   More generally,
such behavior can arise in models that possess one (or more)
off-diagonal squared-mass parameters in addition to the diagonal scalar
squared-masses. When the off-diagonal squared-mass
parameters are taken large [keeping the dimensionless
Higgs self-couplings fixed and $\lsim{\cal O}(1)$], the heavy Higgs
states decouple, while both light and heavy Higgs bosons remain
weakly-coupled.  In this decoupling limit,
exactly one neutral CP-even Higgs
scalar remains light, and its properties are precisely those of the
(weakly-coupled) Standard Model Higgs boson.  That is,
$\hl\simeq\hsm$, with
$\mhl\sim{\cal O}(\mz)$, and all other non-minimal Higgs states are
significantly heavier than $\mhl$.  Squared-mass splittings of the heavy
Higgs
states are of ${\cal O}(\mz^2)$, which means that all heavy Higgs states
are approximately degenerate, with mass differences of order
$\mz^2/\mha$ (where $\mha$ is approximately equal to the common heavy
Higgs mass scale).  In contrast, if the non-minimal Higgs sector is
weakly coupled but far from the decoupling limit, then $\hl$ is not
separated in mass from the other Higgs states.  In this case, the
properties\footnote{The basic property of the Higgs coupling strength
proportional to mass is maintained.  But,
the precise coupling strength patterns of
$\hl$ will differ from those of $\hsm$ in the non-decoupling
limit.}~~of $\hl$ differ significantly from those of $\hsm$.

In the MSSM, the decoupling limit corresponds to the limit of
$\mha\gg\mz$.
In this case, the tree-level Higgs masses take on a particularly
simple form
\beqno
\mhl^2\simeq\ &\mz^2\cos^2 2\beta\,,\nonumber \\[3pt]
\mhh^2\simeq\ &\mha^2+\mz^2\sin^2 2\beta\,,\nonumber \\[3pt]
\mhpm^2=\ & \mha^2+\mw^2\,,\nonumber \\[3pt]
\cos^2(\beta-&\alpha)\simeq\,{\mz^4\sin^2 4\beta\over 4\mha^4}\,.
\label{largema}
\eeqno
A number of consequences are immediately apparent. First,
$\mha\simeq\mhh
\simeq\mhpm$, up to corrections of ${\cal O}(\mz^2/\mha)$.  Second,
$\cos(\beta-\alpha)\simeq 0$ up to corrections of ${\cal O}(\mz^2/\mha^2)$.
One can also check that when $\cosbma=0$, the $\hl$ couplings to vector
boson pairs and fermion pairs, and the $(\hl)^3$ and $(\hl)^4$
self-couplings all reduce to the corresponding Standard Model couplings
of $\hsm$.  In general, the couplings of the non-minimal Higgs
states do not vanish in the decoupling limit, with the notable exception
of the couplings involving at least one vector boson and exactly one
non-minimal Higgs boson [as noted below \eq{littletable}].

Should one expect $\mha\gg\mz$ in the MSSM?
Naively, one might expect the masses of all Higgs sector states to be
roughly of ${\cal O}(v)$.  However, somewhat heavier non-minimal
Higgs states often arise in model building.  In
low-energy supersymmetric models, the mass scale of the non-minimal
Higgs states is controlled by the soft-supersymmetry-breaking
parameter $m_{12}^2$ [see \eq{susymhpm}], which could be as large as
a few TeV. For example, in the minimal supergravity (mSUGRA) model,
one finds that $\mha\gg\mz$ over a
large fraction of the mSUGRA parameter space.\cite{susyrev}

Although the radiative corrections to the Higgs masses
can have a profound effect on the phenomenology,
the overall size of such corrections is never larger than
${\cal O}(\mz)$.  As a result,
the tree-level implications of decoupling
for Higgs masses and couplings remain valid when radiative
corrections are taken into account. This observation is
supported by the results displayed in Fig.~1.  Although the
renormalized value of $\cos(\beta-\alpha)$ is somewhat enhanced
above its tree-level value, it continues to vanish up to corrections
of ${\cal O}(\mz^2/\mha^2)$ in the decoupling limit.

It is not unlikely that the first Higgs state to be discovered will be
experimentally indistinguishable from the Standard Model Higgs boson.
This occurs in many theoretical models that exhibit the decoupling
of heavy scalar states.  As noted above, in the decoupling limit
the lightest Higgs state, $\hl$, is a neutral CP-even scalar
with properties nearly identical to those of the $\hsm$, while the
other Higgs bosons of the non-minimal Higgs sector are heavy (compared
to the $Z$) and are approximately mass-degenerate.  Thus, discovery of
$\hl\simeq\hsm$ may shed little light on the dynamics underlying
electroweak symmetry breaking.
In particular, precision measurements are critical
in order to distinguish between $\hl$ and $\hsm$.
In addition, it is crucial to directly detect
and explore the properties of the non-minimal Higgs bosons.
To accomplish these goals, future colliders
of the highest possible energies and luminosities
are essential.

\section{Higgs Hunting at Future Lepton Colliders}

Higgs hunting at future colliders will consist of three phases.  Phase
one is the initial Higgs boson search in which a Higgs signal is found
and confirmed as evidence for new phenomena not described by
Standard Model background.  Phase two will address the
question: should the signal be identified with Higgs physics?  Finally,
phase three will consist of a detailed probe of the Higgs sector and
precise measurements of Higgs sector observables.

The potential for the discovery of the Higgs boson at future colliders is
well documented.\cite{hehsnow,gunreport,gsw}   Here, I shall briefly
focus on the Higgs searches at a future
(and hopefully, next) $e^+e^-$ linear collider (NLC).  The light
CP-even Higgs boson is detected at the NLC via two processes.
The first involves the extension of the LEP-2 search for
\begin{equation}
e^+e^-\to Z\hl
\label{BJ}
\end{equation}
to higher energies.
In addition, a second process can also be significant:
the (virtual) $W^+W^-$ fusion process\footnote{The corresponding $ZZ$
fusion process, $e^+e^-\to e^+e^-Z^*Z^*\to e^+e^-\hl$ is suppressed
by about a factor of ten relative to the $W^+W^-$ fusion process.
Nevertheless, at large $\sqrt{s}/\mhl$, the $ZZ\to\hl$ fusion rate
compares favorably to that of $e^+e^-\to Z\hl$.  As a result, the $ZZ$
fusion process can be used in some cases to study Higgs properties.}
\begin{equation}
e^+e^- \to\nu\bar\nu W^* W^* \to\nu\bar\nu\hl.
\label{FUSION}
\end{equation}
The fusion cross section grows logarithmically with the center-of-mass
energy and becomes the dominant Higgs production
process at large $\sqrt{s}/\mhl$.
The NLC provides complete
coverage of the MSSM Higgs sector parameter space once the
center-of-mass energy is above 300~GeV.  This conclusion is a
consequence of the MSSM upper bound, $\mhl\lsim 125$~GeV, discussed in
Section~1.  Note that it is possible that the coupling of $\hl$ to
vector boson pairs is suppressed if $\sinbma\sim 0$.  In this case,
$\hl$ will {\it not} be observable in the channels listed above.
However according to Fig.~1,
$\sinbma\ll 1$ occurs only for small values of $\mha$ and
large $\tanb$.  In this case, we note that the $Z\hl\ha$ coupling,
which is proportional to $\cosbma$, is unsuppressed.  In this regard,
the sum rule of \eq{hxi} [with $H=\hl$] is decisive.  Specifically,
the associated production
\begin{equation}
e^+e^-\to\hl\ha
\label{ha}
\end{equation}
provides an addition discovery channel if
$\mha\lsim\sqrt{s}/2$.\footnote{Although $e^+e^-\to\hl\ha$ is
kinematically allowed when $\sqrt{s}/2\lsim\mha\leq\sqrt{s}-\mhl$, the
rate for this process is suppressed (due to the suppression of the
$Z\ha\hl$ coupling in the decoupling limit).}~~If no Higgs signal
is seen in either direct
$\hl Z$ or associated $\hl\ha$ production, then the MSSM can be
unambiguously rule out at the NLC.

Let us suppose that $\hl$ is discovered.  If its properties are
significantly different from $\hsm$, then we know that the Higgs
sector is far from the decoupling limit.  This will be true only if
the non-minimal Higgs sector states ($\ha$, $\hh$, and $\hpm$) are all
rather light, with masses of ${\cal O}(\mz)$.  This would present
a considerable opportunity for the NLC to probe the details of the MSSM
Higgs sector.  In particular, the processes $e^+e^-\to Z\hl$, $Z\hh$,
$\hl\ha$, $\hh\ha$ and $H^+ H^-$ would all be observable at NLC
running at or above its nominal center-of-mass energy of
$\sqrt{s}=500$~GeV.
In contrast, if none of the non-minimal Higgs sector states are
observed at the NLC, then the decoupling limit must be in effect.

The phenomenological consequences of the decoupling regime are both
disappointing and challenging.  In this case, $\hl$ (once discovered)
will exhibit all the expected properties of $\hsm$.
In order to confirm the existence of the non-minimal Higgs sector, one
must either detect a deviation from Standard
Model Higgs physics via precision measurements of $\hl$, or
directly detect the non-minimal Higgs states and explore their
properties.
Clearly, the main obstacle for the discovery of non-minimal Higgs
states at the NLC is the limit of the center-of-mass
energy, which determines the upper limit of the Higgs boson discovery
reach.  In particular, because of the suppressed Higgs boson--gauge
boson couplings in the decoupling limit, the
production of a single heavy non-minimal Higgs state in association with
either $\hl$ or $Z$ is suppressed, even if the process is kinematically
allowed [\eg, recall the footnote below \eq{ha}].
The coupling of the $Z$ to two heavy non-minimal Higgs states
(all approximately equal in mass to $\mha$) is
unsuppressed; however, the corresponding
processes are kinematically forbidden for $\sqrt{s}\lsim 2\mha$.  Thus,
the heavy Higgs states of the MSSM can be produced in
sufficient number and detected only if $\sqrt{s}\gsim 2\mha$.\cite{DECP}  The
discovery reach could in principle be extended by employing the
$\gamma\gamma$ collider mode of the NLC.
In the latter case, the heavy
neutral Higgs states can be produced singly by $s$-channel resonance
production.
In this mode of operation, the search for $\gamma\gamma\to\ha$ and
$\gamma\gamma\to\hh$ can
somewhat improve the non-minimal Higgs mass discovery
reach of the NLC.\cite{gunhab}

\section{Higgs Boson Production at an \boldmath$e^-e^-$ Collider}

Among all the NLC Higgs production mechanisms
cited in Section~3, only $ZZ$-fusion is applicable for Higgs production
at an $e^-e^-$ collider.  The
leading single Higgs production mechanism at an $e^-e^-$
collider is:
\begin{equation}
e^-e^- \to e^-e^- Z^* Z^* \to e^-e^- H\qquad (H=\hh,\hl)\,,
\label{zzfusion}
\end{equation}
where $Z^*$ is a virtual $Z$ boson.\footnote{A second
process, $e^+e^- \to e^-e^- \gamma^* \gamma^* \to e^-e^-H$
($H=\hh,\hl,$ and $\ha$), can also produce the CP-odd Higgs boson
($\ha$).\cite{gangofnine} However, this is a one-loop process and is
thus suppressed in rate as compared to the
$ZZ$-fusion production of the CP-even Higgs bosons.}~~Cross-sections
for the production of $\hsm$ via $ZZ$-fusion have been
obtained in \Ref{han}.  To obtain the corresponding cross sections
for the production of $\hl$ [$\hh$], the
results of \Ref{han} should be multiplied by an overall factor of
$\sintwobma$ [$\costwobma$]. The $ZZ$-fusion rates are quite small; for
$\sqrt{s}=500$~GeV and $\mhl=100$~GeV [with, \eg, $\sinbma=1$ in the
decoupling limit], \Ref{han} quotes $\sigma(e^-e^- \to e^-e^-
\hl)\simeq 9$~fb.  The corresponding rates for $e^+e^-\to Z\hl$
at the NLC are an order of magnitude larger.  For larger
Higgs masses, the $ZZ$-fusion rates become significantly smaller.
However, as noted in the footnote preceding \eq{FUSION},
the rate for $ZZ$-fusion can
become appreciable in size relative to $e^+e^-\to Z\hl$ for large
$\sqrt{s}/\mhl$.  If the Higgs boson decays invisibly,\footnote{This
is unlikely in the MSSM except at unusual points of the parameter space
where the Higgs boson decays dominantly to a pair of the lightest
supersymmetric particles.}~~then both processes provide the possibility
of indirect Higgs detection by measuring the invariant mass of the
system that recoils against the Higgs boson.\cite{pmink}

Let us now consider the possibility of double Higgs production at an
$e^-e^-$ collider.  Double Higgs production can only occur via gauge
boson fusion, and results in four body final states: $e^-e^-\to
\ell\ell HH$, where $H$ is a charged or neutral Higgs boson and $\ell$
is either an electron or $\nu_e$.  Cross-sections for these
processes are known to be quite small.
Nevertheless, if any of these processes could be detected, it could
provide a very sensitive test of the underlying Higgs boson dynamics.
In particular, such processes can probe some of the triple Higgs
couplings, which would provide a direct measure of the Higgs potential.
In the context of the MSSM, one could perhaps verify that the
Higgs self-couplings are given by gauge couplings as discussed in
Section~1.  Unfortunately, there are only a
few cases where the measurement of Higgs self-couplings at the NLC has
been shown to be viable.\cite{DHZ}   In particular, the four-body
phase space leads to a substantial suppression
that could only be overcome by machines of the highest possible
energies and luminosities.

The corresponding calculations for double Higgs production at $e^-e^-$
colliders have not yet been carried out systematically.  An initial
study was reported in \Ref{eminusrizzo}.  Here, I shall indicate
the complete set of double Higgs processes that can be produced in
$e^-e^-$ collisions.  Detailed calculations will not be provided here,
although general expectations based on the decoupling limit will be
indicated.
The mechanism for double Higgs production is vector boson fusion
(symbolically indicated by $VV\to HH$).
Each electron can emit either a virtual $W^-$, $Z$ or photon.  In
Table~1, I list the possible fusion mechanisms and the corresponding
Higgs bosons that can appear in the final state.

The last column of Table 1 indicates whether the process is
suppressed in the decoupling limit.  The case of $H^- H^-$ fusion is
particularly noteworthy.  As shown in \Ref{eminusrizzo}, there are
a number of contributions to the amplitude for $W^- W^-\to H^-H^-$,
corresponding to the $t$-channel and $u$-channel exchange of $\hl$,
$\hh$ and $\ha$.  In the decoupling limit (where $\mhh\simeq \mha$),
the $t$-channel amplitude is proportional to
\beq\label{hminushminus}
{\costwobma\over t-\mhl^2}+{\sintwobma\over t-\mhh^2}-{1\over
t-\mha^2}\simeq\costwobma\left[{1\over t-\mhl^2}- {1\over
t-\mha^2}\right]\,.\eeq
The $u$-channel amplitude behaves similarly (\ie, simply interchange $t$
and $u$).  Thus, in the decoupling limit, where $\costwobma\ll 1$, we
see that double charged Higgs production is suppressed.

\begin{table}
\begin{center}
\tcaption{$VV \to HH$ Processes}
\vskip6pt
\begin{tabular}{llc}
Fusion& Final States& Decoupling? \\[3pt]
\hline  \noalign{\vskip3pt}
$W^-W^-$& \quad$H^-H^-$& yes \\[6pt]
$\gamma W^-,ZW^-$& $\left\{ \matrix{ h^0H^- \qquad\quad\cr
H^0H^-,A^0H^- } \right.$&
         $\matrix{ {\rm yes} \cr \noalign{\vskip3pt}{\rm no}}$\\[1pc]
$\gamma\gamma,ZZ$& \quad$ H^+H^-$& no \\[6pt]
$ZZ$&  $\left\{ \matrix{ h^0A^0 \qquad\qquad\qquad\cr
h^0h^0,H^0H^0,A^0A^0} \right.$&
      $\matrix{ {\rm yes} \cr\noalign{\vskip3pt} {\rm no}}$  \\[1pc]
\hline
\end{tabular}
\end{center}
\end{table}

The above considerations are based on a tree-level analysis and are
confirmed in \Ref{eminusrizzo}.  However, \Ref{eminusrizzo} goes on to
argue that the cancelation illustrated in \eq{hminushminus} is
significantly softened when radiative corrections are taken into
account.
In a first approximation, one can account for the radiative corrections
by computing the radiatively corrected CP-even Higgs squared-mass
matrix, and extracting the the corresponding CP-even Higgs mixing angle
$\alpha$, as discussed at the end of Section~1.  The
radiatively-corrected value of
$\costwobma$ (shown in Fig.~1) would then be employed in
\eq{hminushminus}.  This procedure would result in a slight enhancement
of the $H^-H^-$ production cross section due to a slightly larger value
for the radiatively-corrected $\costwobma$.  However, the resulting
cross section would still exhibit the decoupling limit suppression in
apparent conflict with the results of \Ref{eminusrizzo}.  In this case,
a better computation is needed, in which genuine radiative corrections
to the full scattering amplitude are also calculated.  It is possible
that one could find new terms at one-loop which do not vanish in the
decoupling limit.\footnote{However, such terms cannot be logarithmically
enhanced.}~~Whether such terms lead to a significant enhancement of the
$H^-H^-$ production amplitude remains to be checked.

Nevertheless, I believe that the behavior of the tree-level amplitudes
in the decoupling limit does provide a rough guide to the likely size of
the double Higgs production rates.
It should be
noted that the $ZZ$ fusion processes can involve $s$-channel
Higgs exchange; thus these processes are potentially sensitive to triple
Higgs couplings.  Finally, only tree-level $\gamma\gamma$ fusion
processes are listed in Table~1.  With the exception of a $\gamma\gamma$
collider, the $\gamma\gamma$ fusion processes through a loop to a pair
of Higgs bosons is suppressed relative to the tree-level processes of
Table~1.

For completeness, one can also consider Higgs boson production in
association with vector bosons via gauge boson fusion (indicated
symbolically by $VV\to VH$).  Again, these processes are severely
suppressed by the four-body phase space.  Additional suppressions in the
decoupling limit can also occur, as exhibited in Table~2.

\begin{table}[h]
\begin{center}
\tcaption{$VV \to VH$ Processes}
\vskip6pt
\begin{tabular}{llc}
Fusion& Final States& Decoupling? \\[3pt]
\hline  \noalign{\vskip3pt}
$W^-W^-$&\quad $W^-H^-$& yes \\[6pt]
$\gamma W^-$& $\left\{ \matrix{ W^-h^0 \;\cr  W^-H^0 } \right.$&
         $\matrix{ {\rm no} \cr \noalign{\vskip3pt}{\rm yes}}$\\[1pc]
$ZW^-$& $\left\{ \matrix{ W^-h^0 \qquad\quad\cr  W^-H^0, ZH^-}\right.$&
         $\matrix{ {\rm no} \cr \noalign{\vskip3pt} {\rm yes}}$\\[1pc]
$ZZ$&  $\left\{ \matrix{ ZH^0, ZA^0 \cr   Zh^0\quad\qquad\!} \right.$&
      $\matrix{ {\rm yes} \cr\noalign{\vskip3pt} {\rm no}}$  \\[1pc]
\hline
\end{tabular}
\end{center}
\end{table}


Of course, if all the non-minimal Higgs states
have mass of ${\cal O}(\mz)$, indicating a substantial deviation from
the decoupling limit, then the attendant factors of $\costwobma$ in the
rates for $HH$ and $HV$ production will not result in a significant
suppression.
In this case, the ``decoupling'' columns of Tables 1 and 2 are no longer
relevant.  Moreover, the relatively light final state Higgs masses
could allow for a study of $HH$ and $HV$ production at a very
high energy $e^-e^-$ collider of sufficient luminosity.  The
computations of \Ref{eminusrizzo} indicate the possibility of
cross sections as large as 0.1~fb for $\sqrt{s}$ in the range of 1 to 2
TeV.  Luminosities of the $e^-e^-$ collider in excess of
$10^{34}$~cm$^{-2}$~sec$^{-1}$ would be required.

For probing the MSSM Higgs sector, the $e^-e^-$ collider provides little
advantage over an $e^+e^-$ or $\gamma\gamma$ collider.  Rates for Higgs
production tend to be very small, and would require machines of the
highest possible energy and luminosity.  The true motivation for an
$e^-e^-$ collider must lie in more exotic processes.  \Ref{jfgexotic}
discusses the potential of an $e^-e^-$ collider to probe a variety of
exotic Higgs sectors.  It remains to be seen whether the
simplest Higgs sectors of the Standard Model or the MSSM will be
confirmed, or whether nature chooses a more unexpected
path for the dynamics underlying electroweak symmetry breaking.

\nonumsection{References}

\end{document}